\documentclass[12pt]{iopart}

\usepackage{ifpdf}
\ifpdf
\usepackage{subfigure}
\usepackage[pdftex]{graphicx}
\else
\usepackage{graphicx}
\fi

\usepackage{psfrag}
\usepackage{epstopdf}
\usepackage{graphicx}
\DeclareGraphicsExtensions{.eps}

\expandafter\let\csname equation*\endcsname\relax
\expandafter\let\csname endequation*\endcsname\relax

\usepackage{amsmath,graphicx,epsfig,bm,color}
\usepackage{graphics}
\usepackage{iopams}  

\definecolor{red}{rgb}{1,0,0}
\definecolor{orange}{rgb}{1,0.5,0}
\definecolor{blue}{rgb}{0,0,1}
\definecolor{green}{rgb}{0,1,0}

\begin{document}

\title[Time-dependent RG on an adiabatic random quantum Ising model]{Time-dependent Real-space Renormalization-Group Approach: 
application to an adiabatic random quantum Ising model}
\author {Peter Mason, Alexandre M. Zagoskin and Joseph J. Betouras}
\address {Department of Physics, Loughborough University, Loughborough, 
LE11 3TU, United Kingdom}


\begin{abstract}
We develop a time-dependent real-space renormalization-group approach which can be applied to Hamiltonians with time-dependent random terms. To illustrate the renormalization-group analysis, we focus on the quantum Ising Hamiltonian with random site- and time-dependent (adiabatic) transverse-field and nearest-neighbour exchange couplings. We demonstrate how the method works in detail, by calculating the off-critical flows and recovering the ground state properties of the Hamiltonian such as magnetization and correlation functions. The adiabatic time allows us to traverse the parameter space, remaining near-to the ground state which is broadened if the rate of change of the Hamiltonian is finite. The quantum critical point, or points, depend on time through the time-dependence of the parameters of the Hamiltonian. We, furthermore, make connections with Kibble-Zurek dynamics and provide a scaling argument for the density of defects as we adiabatically pass through the critical point of the system.
\end{abstract}

\address{Corresponding authors; P. Mason@lboro.ac.uk, J. Betouras@lboro.ac.uk.}
\maketitle

\section*{I. Introduction}

An interacting quantum system evolving at zero temperature can demonstrate various forms of approach to equilibrium even with no loss of phase coherence. In the past, detailed experimental \cite{Sadler} and theoretical studies \cite{Gritsev, Lamacraft, Lauchli, Flesch, Rossini, Faribault} for systems prepared by a quantum quench across a phase transition have studied this kind of out-of-equilibrium path. A system is prepared in the ground state for certain values of parameters, which are then rapidly changed to values for which the ground state (GS) is known to be in a different phase. Experiments on ultracold atoms are particularly useful to probe this physics because they are essentially closed quantum systems on rather long time scales compared to the basic dynamical time scales of the system. As such, fundamental questions as to whether many physical properties equilibrate after the quench exponentially in time and the system thermalizes (so it can be described by an effective temperature) can be addressed. 

Apart from instant quenches of an interacting quantum system, there is much attention on those Hamiltonians which change smoothly across a quantum phase transition \cite{dz1,dz2,pol,green,kz}. In general a single tunable parameter, or perturbation, is chosen to address this transition, and the non-equilibrium dynamics such as the growth of density defects or entanglement entropy is then studied as a function of this parameter. For a second-order transition critical point, various averaged physical quantities such as the excitation density and energy show power laws as the rate of change approaches zero, with exponents determined by the universal physics of the quantum critical point. The system evolves after such a sweep to a steady state for some quantities, but its energy distribution remains nonthermal \cite{green}. 

Quantum spin chains provide prime examples of the interacting quantum systems where the evolution of out-of-equilibrium many-body systems \cite{eisert}, that possess many degrees of freedom, is theoretically illustrated. The addition of disorder in the system provides another dimension which needs to be addressed. The properties of quantum spin chains with quenched randomness at low temperatures have been the subject of interest for a number of years. In the case of a one-dimensional transverse-field quantum Ising model with randomness, a renormalization-group (RG) treatment shows that the disorder grows without bound \cite{fisher,fisher92,fisher94} - the flow is towards an infinite-randomness fixed point. This strong-disorder renormalization-group approach has seen further successful progress and development to tackle issues related to many-body localisation in interacting models with disorder \cite{va_rev,altman_vosk}, such as establishing the form of the growth of the entropy in an XXZ spin-chain \cite{vosk_altman,xxz}, as well as establishing the physics around the phase transition between the many-body localised phase and the thermal phase for a general interacting model \cite{potter,pekker,v_huse_a,zhang,dumit}, and investigating Floquet dynamics in periodically driven random quantum spin chains \cite{monthus1,monthus2}.

In the present work, we address the problem of a random quantum spin chain, requiring control of more than one parameter as we pass through a quantum critical point. We will as such concentrate on an adiabatic quantum Hamiltonian. This is also an important question in the field of adiabatic quantum computation, with a time-dependent Hamiltonian of the generic form
\begin{equation}
	\label{gov_ham}
	H(t)=H_{\text{initial}}(t)+H_{\text{final}}(t),
\end{equation}
subject to initial conditions that $H_{\text{final}}(0)=0$ and final conditions that $H_{\text{initial}}(T)=0$, where $T$ is the end-point of the computation. The ground, and easily reachable, state of the initial Hamiltonian, $H_{\text{initial}}$, is assumed known, whereas the ground state of the final Hamiltonian at $t=T$, $H_{\text{final}}$, is unknown and encodes the desired solution to the computation. The adiabatic evolution 
of the Hamiltonian is such that the computation always remains in, or close to, the local ground state at some given instant in time. Crucially, the computation will therefore terminate (ideally in) the ground state of $H_{\text{final}}$.

The inclusion of adiabatic parameters means that the energy gap between the ground state and first excited state is now necessarily time-dependent, such that it can vary depending on the value of time (i.e.\ because of the quantities $H_{\text{initial}}(t)$ and $H_{\text{final}}(t)$) in the computation.
As such, remaining in the ground state, or removing any excited states, requires the adiabatic process to progress sufficiently slowly. However, if the quantum Hamiltonian exhibits a quantum critical point, the energy gap will, at this point, decrease to zero, and the computation will lose adiabaticity. For the quantum computation therefore, the transition through the quantum critical point will lead to the development of excited states, or defects, the density of which depend on the rate of transition.

We address the issue of a generic time dependence in the quantum Hamiltonian (Eq.\ (\ref{gov_ham})), focussing on developing a complete analysis of the state of the system at any given time,
as well as the dynamics across a given quantum critical point. The time-dependence (as we will see) provides a tool to shift in time the location of the quantum critical point. It also allows us, through the choice of the adiabatic parameters, to increase (to greater than one) the number of critical points. This can be particularly important in the development of a device to simulate the quantum critical point: Optimisation problems commonly in the NP-complete and NP-hard classes can be mapped onto an Ising model \cite{comp,das_rmp} (indeed the architecture of, for example the D:Wave machines \cite{dwave1,dwave2}, is designed to use a quantum adiabatic protocol, and run based on an adiabatic quantum Ising Hamiltonian). As such we will concentrate our analysis on an adiabatic Ising Hamiltonian with random transverse-fields and exchange couplings, but simplify the setup to consider an infinitely long one-dimensional chain of Ising spins. Such 1D spin chains governed by a quantum Ising Hamiltonian can be experimentally realised as flux qubits; see for instance \cite{az_2,az_1}.

The strong-disorder RG approach that we develop is the time-dependent extension of that formulated in detail for the time-independent transverse-field Ising Hamiltonian by D.\ S.\ Fisher in a series of papers in the 1990's \cite{fisher,fisher92,fisher94}. The work was based on the perturbative technique developed by Dasgupta and Ma \cite{dm} that safely allows for the systematic removal of the high-energy degrees of freedom on the spin-$1/2$ Heisenberg chain. The approach that we will employ, renormalising the bond and exchange couplings at any given instant in time, will provide access to the (local in time) properties, such as the magnetisation or correlation lengths of the system state at any given time, be that in the paramagnetic state or the ferromagnetic state. We will also be able to analyse the off-critical flows around the quantum critical point and to establish the conditions for a time-dependent duality across this critical point. This analysis of the state properties enables us to investigate the non-equilibrium dynamics which follow the Kibble-Zurek mechanism \cite{dz1,dz2,pol,italians,italians2,kolo} as the governing Hamiltonian (Eq.\ (\ref{gov_ham})) is transitioned adiabatically through a critical point. We are then able to provide a characteristic rate of adiabatic transition across the critical point and link this rate to a cut-off energy scale related to the validity of the RG scheme.

The paper is organised as follows: in Sect.\ II we outline the main steps in the renormalization-group approach, focussing on the one-dimensional random transverse-field Ising model and the (adiabatic) time conditions under which this approach is valid. We are able to establish the forms of the distribution functions for the transverse-field and exchange couplings. These allow us to identify a measure of the `distance' - as a function of the random couplings and adiabatic parameters - from the quantum critical point separating the paramagnetic state from the ferromagnetic state. Section III identifies a critical time scale based on the adiabatic coupling parameters and uses a scaling argument to provide an estimate for the density of defects as we transition through the quantum critical point. We conclude with a discussion in Sect.\ IV and include some further calculations in the Appendix.

\section*{II. The Real-Space Renormalization-Group 
Approach}
In this section we present the analysis of the adiabatic Ising Hamiltonian in the strong-disorder limit. We concentrate on a renormalization-group (RG) approach and discuss the role that time plays as well as the underlying conditions of the analysis. Then, we proceed by writing down the time-dependent coupled integro-differential RG flow equations (in terms of the distributions functions of the exchange and site couplings) that govern the distribution of the random variables in our model. These flow equations are then solved explicitly.
\subsection*{A. The Adiabatic  Random Quantum Ising 
Hamiltonian}

To illustrate our approach to the development of a time-dependent strong-disorder RG scheme, we study the nearest neighbour time-dependent random transverse-field Ising Hamiltonian
\begin{equation}
	\label{ham}
	H(t)=-\sum_{i=1}^{N}h_i(t)\sigma_i^x-\sum_{<i,j>}^NJ_{ij}(t)\sigma_i^z\sigma_j^z
\end{equation}
on the one-dimensional infinite chain, where the parameters $h_i(t)$ and $J_{ij}(t)$ are time-dependent random variables and denote the transverse fields and the nearest-neighbour interactions, respectively. To study the adiabatic evolution of the fixed point distributions of the above Hamiltonian the following boundary conditions are imposed:
\begin{equation}
	\label{bc}
	J_{ij}(0)=0,\quad h_i(T)=0,
\end{equation}
so that the evolution runs for times $t\in[0,T]$, with beginning- and end-point Hamiltonians given by
\begin{subequations}
	\begin{eqnarray}
		H_{\text{initial}}\equiv H(t=0)&=&-\sum_i^N h_i(0)\sigma_i^x,\\
		H_{\text{final}}\equiv H(t=T)&=&-\sum_{<i,j>}^N J_{ij}(T)\sigma_i^z\sigma_j^z,
	\end{eqnarray}
\end{subequations}
respectively. 

The governing Hamiltonian, Equation (\ref{ham}), is termed the {\it{adiabatic}} random quantum Ising model (ARQIM). This is to be contrasted to the {\it{non-adiabatic}} random quantum Ising Hamiltonian, considered by, amongst others, Fisher \cite{fisher}, that  can be thought of as a special case of the model studied here, and occurs when $h_i(t)$ and $J_{ij}(t)$ are no longer functions of time.
A further special case of the ARQIM occurs when $t=T$, at which point we obtain the classical Ising model (that is identical to the end-point Hamiltonian, $H_{\text{final}}$, given above).

A transformation can be performed on the ARQIM  by taking 
\begin{equation}
	s_i^z=\prod_{j\le i}\sigma_j^x,\qquad
	s_i^x=\sigma_i^z\sigma_{i+1}^z,
\end{equation}
such that, on exchange of the $h_i(t)$ and $J_{ij}(t)$, the ARQIM is recovered again: this identifies a duality transformation 
$h_i(t)\longleftrightarrow J_{ij}(t)$,
a feature that will be used in the RG analysis later in this section. In analogy with the non-adiabatic Hamiltonian, subject to certain conditions that we discuss below, the ARQIM contains a quantum critical point which relates the magnitudes of the site and bond couplings, $h_i(t)$ or $J_{ij}(t)$. For now, there are no restrictions on these relative magnitudes, however at a later stage, the limit around, and the dynamics across, the critical point $\delta=0$, will be studied. An expression for $\delta$ has been obtained in the non-adiabatic case in \cite{fisher}; we modify this expression to include the adiabatic parameters, to get
\begin{equation}
	\delta\approx\frac{\Delta_{h(t)}-\Delta_{J(t)}}{\text{Var}(\log({J(t)}))+\text{Var}(\log({h(t)}))},
	\label{delta}
\end{equation}
where $\Delta_{h(t)}=\overline{\log({{h(t)}})}$, $\Delta_{J(t)}=\overline{\log({{J(t)}})}$, and where ${J(t)}$ and ${h(t)}$ are the sets that contain the ${J_{ij}(t)}$ and ${h_i(t)}$. The critical point is at $\delta=0$ (note that our analysis includes the possibility of greater than one critical point in the time internal $[0,T]$), which occurs when $\Delta_{h(t)}=\Delta_{J(t)}$. If $\delta>0$ then the ground state under the renormalization procedure will be paramagnetic, whereas in the other limit, $\delta<0$, the ground state will be ferromagnetic. Either side of the critical point rare regions effects are important; these Griffiths phases persist well into the paramagnetic or ferromagnetic regions, however we note that our boundary conditions (Eq.\ (\ref{bc})) impose that at some $\delta$ both paramagnetic and ferromagnetic rare region effects are lost \cite{igloi1,igloi2,vojta}.

\subsection*{B. The Role of Time}

At this stage it is appropriate to discuss the role of time in the renormalization procedure. We will allow for an explicit time dependence in the distributions of the $J(t)'s$ and $h(t)'s$.
In general, the renormalization leads to the low-energy fixed point ground state solution, that is either paramagnetic ($\delta>0$) or ferromagnetic ($\delta<0$). Here, time plays the role of a weighting on the random couplings, thus influencing the form of the ground state. However, we must also consider the time-flow as being in competition with the RG flow. As such, at each given time, the RG flow will (in general) not lead to the ground state, rather an excited energy state. We can look at the time perturbation (essentially a different weighting on the random couplings) as a perturbation away from the ground state: a purely RG flow would allow this perturbation to relax to the (new) ground state. However, we do not allow the RG flow to do this, instead we assume that the time perturbation will leave us in an excited state. Since we are only interested in adiabatic time flow, we expect that the excited states that we recover are close to the ground state.


By writing $h_i(t)=B_i(t)\tilde{h}_i$ and $J_{ij}(t)=A_{ij}(t)\tilde{J}_{ij}$, the time-dependence of $h_i(t)$ and $J_{ij}(t)$ can be separated out. Then $\tilde{h}_i$ and $\tilde{J}_{ij}$ are strictly time-independent independent random distributions, and $A_{ij}(t)$ and $B_i(t)$ are (adiabatic) bond/site-dependent non-random, hence controllable, functions. The form of the time-dependent functions $A_{ij}(t)$ and $B_i(t)$ are, at this stage, left entirely general and independent, and for the majority of our analysis we will work with the functions $h_i(t)$ and $J_{ij}(t)$, i.e.\ we will not assume that the time dependence can be separated out from the random variables. However, in order to clarify the RG procedure or the physics, we will at various stages separate out the time dependence. 

\begin{figure}
\begin{center}
\includegraphics[scale=0.3]{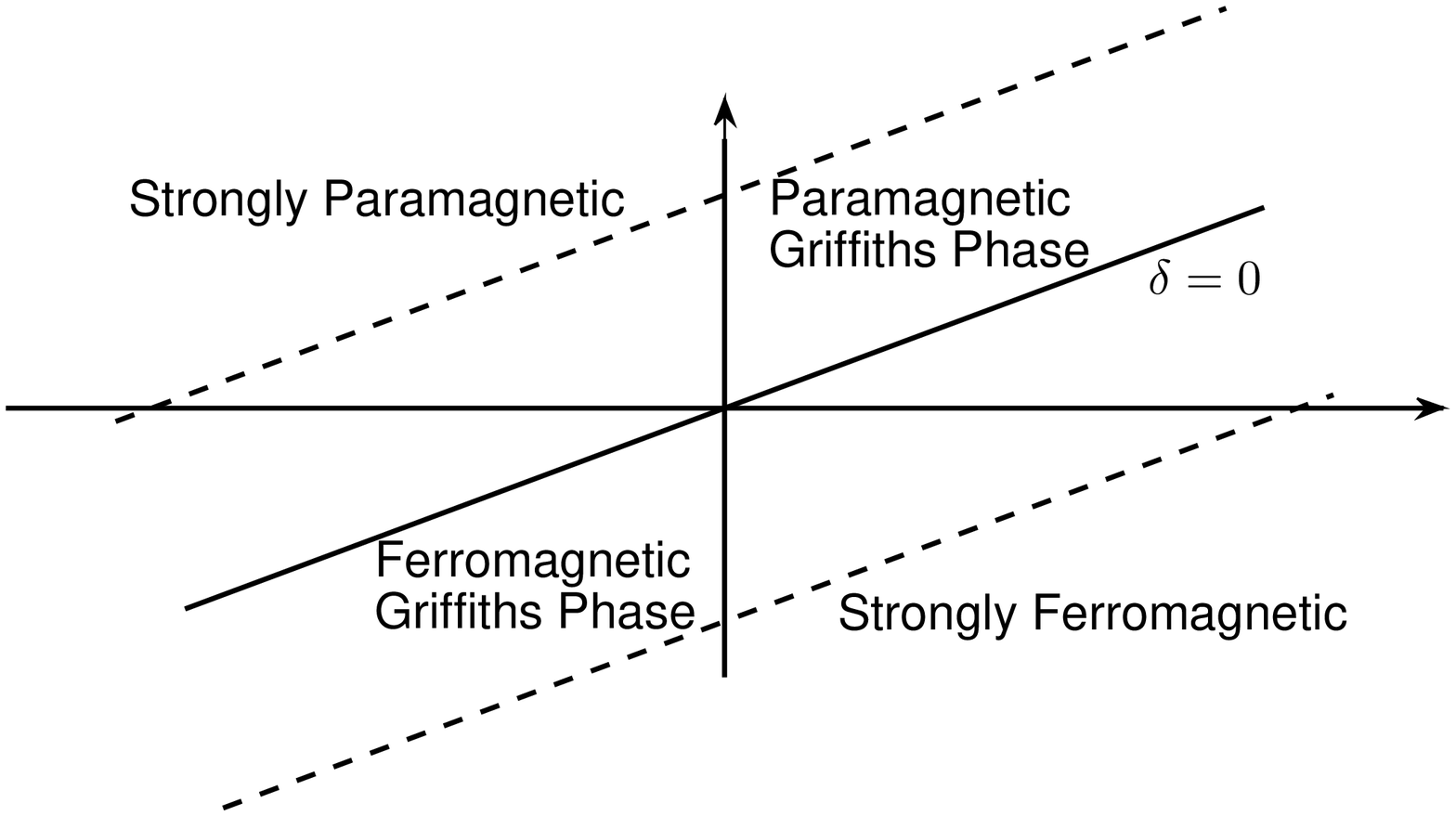}
\end{center}
\begin{picture}(0,0)(10,10)
\put(320,88) {\tiny{$\Delta_t$}}
\put(200,132) {\tiny{$\Delta_{\tilde{h}}-\Delta_{\tilde{J}}$}}
\end{picture}
\vspace{-0.7cm}
\caption{The critical line, $\delta=0$ separating the paramagnetic ($\delta>0$) from the ferromagnetic ($\delta<0$) region.}
\label{critical_line}
\end{figure}

In the case that we can separate out the time-dependence, Eq.\ (\ref{delta}) can be written as
\begin{equation}
	\delta\approx\frac{\Delta_{\tilde{h}}-\Delta_{\tilde{J}}-\Delta_t}{\text{Var}(\log(\tilde{J}))+\text{Var}(\log(\tilde{h}))},
	\label{delta_time}
\end{equation}
where $\Delta_{\tilde{h}}=\overline{\log({{\tilde{h}}})}$, $\Delta_{\tilde{J}}=\overline{\log({{\tilde{J}}})}$ and $\Delta_t={\log({A}(t)/{B}(t))}$ (we note that we consider a disorder average and that $A_{ij}(t)=A(t)$ and $B_{i}(t)=B(t)$ are non-random functions of time as will be considered explicitly below), and $\tilde{J}$, $\tilde{h}$, are the sets that contain the $\tilde{J}_{ij}$, $\tilde{h}_i$ respectively.
In particular, $\Delta_t$ can take positive or negative values (note that $\Delta_t\in[-\infty,\infty]$). The inclusion of a time dependence in the Hamiltonian gives rise to the $\Delta_t$ parameter, the value of which (either positive or negative) can influence the ground state properties of the system, and leads to the possibility of multiple, independent, critical points (dependent on the values of $A_{ij}(t)$ and $B_{i}(t)$). Schematically, the critical line as a function of $\Delta_{\tilde{h}}-\Delta_{\tilde{J}}$ and $\Delta_t$ is presented in Fig. \ref{critical_line}.

\begin{figure}
\begin{center}
\includegraphics[scale=0.21]{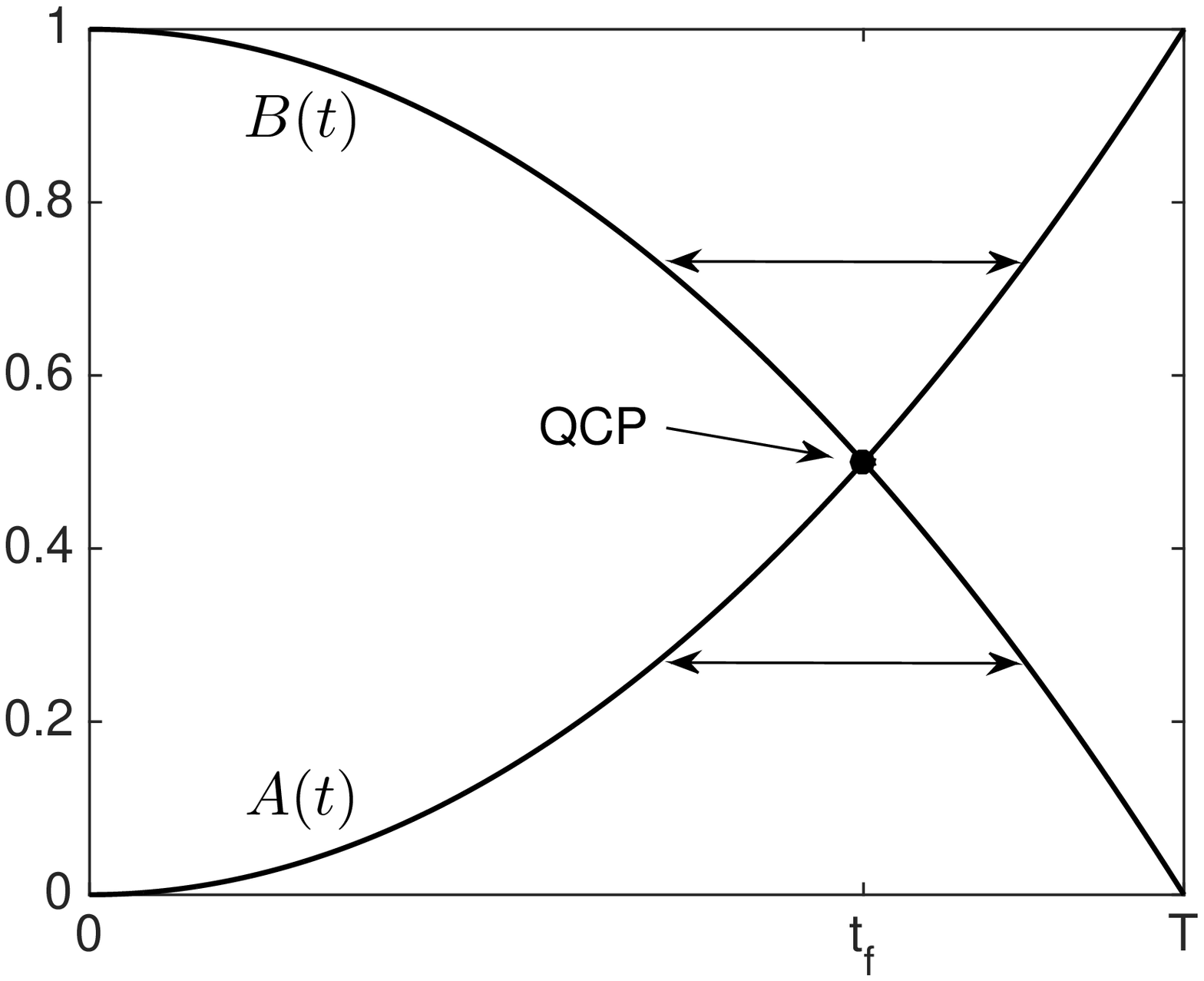}
\includegraphics[scale=0.21]{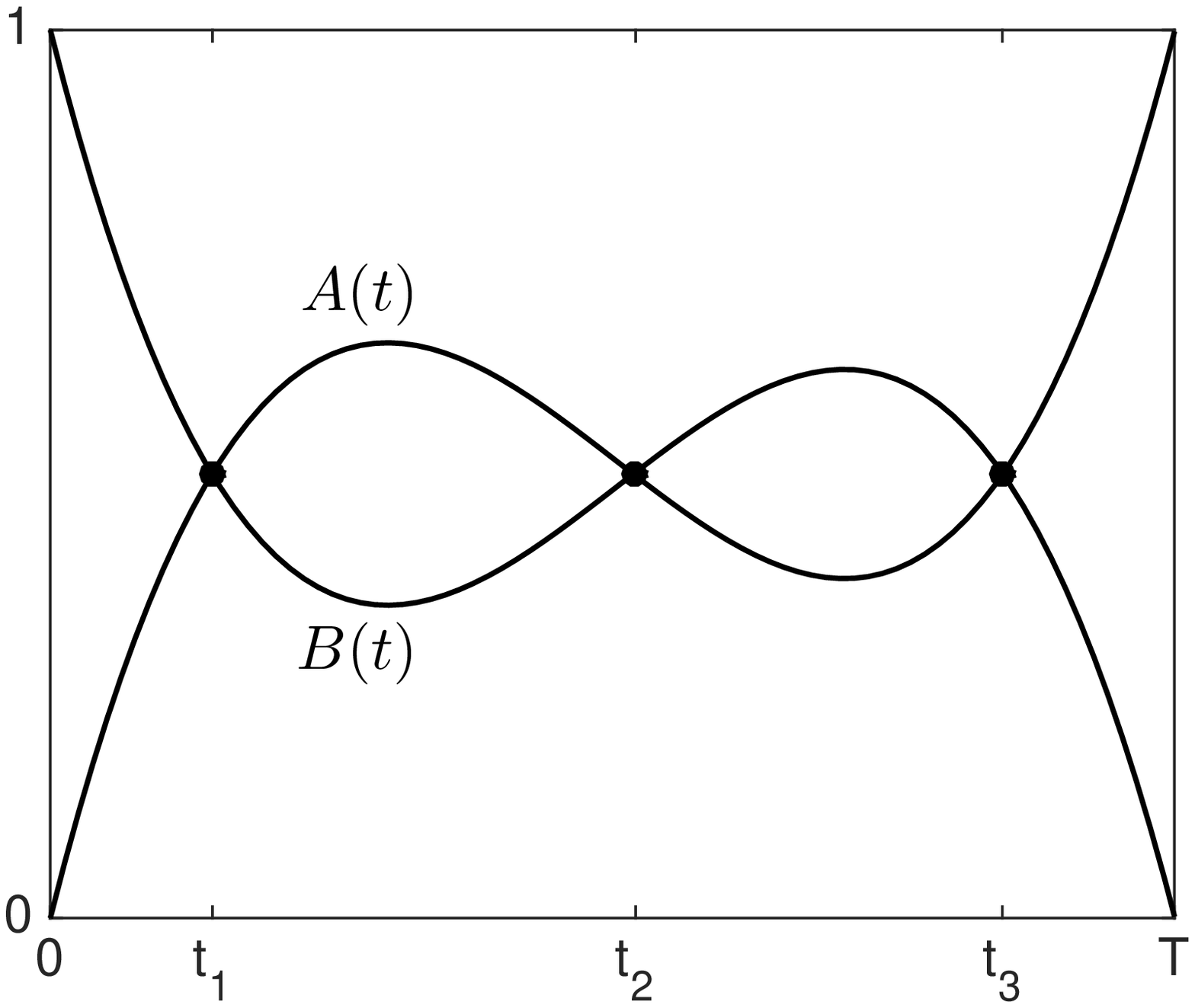}
\end{center}
\begin{picture}(0,0)(10,10)
\put(170,31) {\tiny{$t$}}
\put(290,31) {\tiny{$t$}}
\end{picture}
\vspace{-0.7cm}
\caption{Schematic time-flow under the supposition that $\Delta_{\tilde{h}}=\Delta_{\tilde{J}}$ throughout. (a)  $A(t)=A_{ij}(t)$ and $B(t)=B_i(t)$ $\forall i,j$. There is a quantum critical point (QCP) which occurs when $t=t_f$ and $\delta=0$, here indicated at the point when $A(t_f)=B(t_f)$. The ground state is paramagnetic for $t<t_f$ and ferromagnetic for $t>t_f$. A duality in time exists across the critical point, as indicated by the double sided arrows. (b) The $A_{ij}(t)$ and $B_{i}(t)$ are taken to be in general distinct. There exist multiple (here three) quantum critical points when $\Delta_t=0$, at $t=t_1$, $t_2$ and $t_3$.}
\label{schem}
\end{figure}

In Fig.\ \ref{schem} we give two schematic viewpoints of the time-flow under the supposition that $\Delta_{\tilde{h}}=\Delta_{\tilde{J}}$ throughout. Figure \ref{schem}(a) further supposes that $A(t)=A_{ij}(t)$ and $B(t)=B_i(t)$ $\forall i,j$ (here we choose $A(t)=1-B(t)=(t/T)^2$, but in general they can take any form). There then exists a single quantum critical point at $t=t_f$, where $A(t)=B(t)$, such that for $\delta>0$ (equivalently $t<t_f$) the ground state is paramagnetic, whereas for $\delta<0$ (equivalently $t>t_f$) the ground state is ferromagnetic. In contrast Fig.\ \ref{schem}(b) takes the $A_{ij}(t)$ and $B_{i}(t)$ to be in general distinct, and shows the existence of multiple independent quantum critical points, all located at $\Delta_t=0$. We note that if $\Delta_{\tilde{h}}\neq\Delta_{\tilde{J}}$, then a quantum critical point can still exist: for instance if $\Delta_{\tilde{h}}=\tilde{\alpha}\Delta_{\tilde{J}}$, for some constant $\tilde{\alpha}$, then the quantum critical point exists when $\Delta_t=(\tilde{\alpha}-1)\Delta_{\tilde{J}}$, or equivalently $A(t)=B(t)\exp((\tilde{\alpha}-1)\Delta_{\tilde{J}})$. 

\subsection*{C. Renormalization-Group Formulation} 

The analysis that we formulate in this paper for the ARQIM will concentrate on a strong-disorder RG approach on the joint distribution functions $p(t,J(t),l^S(t);\Omega(t))$ and $r(t,h(t),l^B(t);\Omega(t))$, where $l^B(t)$ is the bond length scale and $l^S(t)$ is the site (spin cluster) length scale. In the RG approach, an energy scale $\Omega(t)$ is introduced \cite{fisher,va_rev}, defined as the maximum of the bond and coupling strengths, i.e.\ $\Omega(t)=\max_{ij}\left(J_{ij}(t),h_i(t)\right)$. The RG analysis proceeds by perturbatively removing the high-energy degrees of freedom, thus reducing the energy scale. 
The RG flows will thus provide the appropriate low-energy physical quantities of interest; after many RG steps the effective distribution functions read $\hat{p}(t,\hat{J}(t),\hat{l}^S(t);\Omega(t))$ and $\hat{r}(t,\hat{h}(t),\hat{l}^B(t);\Omega(t))$, with the `hat' notation indicating that renormalization has taken place. 

Specifically, there are two cases to consider when analysing the governing Hamiltonian: $\Omega(t)$ takes either an element from $J(t)$ or an element from $h(t)$. At this stage a rescaling \cite{fisher,va_rev} of the random variables is made, through the introduction of the following time-dependent functions
\begin{subequations}
	\label{rescalings}
\begin{eqnarray}
	\Gamma(t)&=&\ln\left(\frac{\Omega_I(t)}{\Omega(t)}\right),\\
	\zeta(t)&=&\ln\left(\frac{\Omega(t)}{\hat{J}(t)}\right),\\
	\beta(t)&=&\ln\left(\frac{\Omega(t)}{\hat{h}(t)}\right),
\end{eqnarray}
\end{subequations}
with $\Omega_I(t)$ defined as the initial maximum bond or coupling strength. After each renormalization step we will, in general, find $\Omega(t)<\Omega_I(t)$, so that $\Gamma(t)$ will become a large parameter in the problem, representing the decrease in the energy scale. Note that $\zeta(t)$ and $\beta(t)$ are strictly non-negative, defined on the interval $[0,\infty)$.

In the case of the problem considered in this work, the rescaled joint distribution functions $\hat{p}(t,\hat{J}(t),\hat{l}(t);\Omega(t))$ and $\hat{r}(t,\hat{h}(t),\hat{l}(t);\Omega(t))$ become $\hat{P}(t,\hat{\zeta}(t),\hat{l}(t);\Gamma(t))$ and $\hat{R}(t,\hat{\beta}(t),\hat{l}(t);\Gamma(t))$. 
We wish to find the form of these distribution functions for all time. Given that the RG perturbation steps remain valid (see appendix), we can write down coupled master equations that describe the renormalization flow. To begin, we note that we are
interested in the adiabatic limit of time evolution. As such we can consider a perturbation in time to time-independent master equations. The solution to these equations provides the form of the distribution functions for the off-critical flow. The integro-differential coupled flow equations were originally written down by Fisher \cite{fisher,va_rev}. To form these equations, we must combine two effects. The first is the change in either $\zeta$ or $\beta$ as a result of the renormalisation that causes $\Gamma$ to increase (equivalently $\Omega$ to decrease). The second effect comes from the decimation of either a bond or site and the recombining of a new site or bond. Putting these together, along with a normalisation term gives us
\begin{subequations}
	\label{fishers}
	\begin{eqnarray}
		\hat{P}_{\Gamma}-\hat{P}_{\zeta}-\hat{R}_0\hat{P}* \hat{P}-\left(\hat{P}_0-\hat{R}_0\right)\hat{P}&=&0,\\
		\hat{R}_{\Gamma}-\hat{R}_{\beta}-\hat{P}_0\hat{R}* \hat{R}-\left(\hat{R}_0-\hat{P}_0\right)\hat{R}&=&0,			
	\end{eqnarray}
\end{subequations}
where subscripts $\Gamma$, $\zeta$ and $\beta$ indicate partial differentiation with respect to that variable, `$*$' indicates a convolution, and $\hat{P}_0\equiv\hat{P}|_{\zeta(t)=0}$ and $\hat{R}_0\equiv\hat{R}|_{\beta(t)=0}$.

Now we denote
\begin{subequations}
\begin{eqnarray}
\mathcal{P}\equiv \hat{P}_{\Gamma}-\hat{P}_{\zeta}-\hat{R}_0\hat{P}* \hat{P}-\left(\hat{P}_0-\hat{R}_0\right)\hat{P},\\
\mathcal{R}\equiv \hat{R}_{\Gamma}-\hat{R}_{\beta}-\hat{P}_0\hat{R}* \hat{R}-\left(\hat{R}_0-\hat{P}_0\right)\hat{R},
\end{eqnarray}
\end{subequations}
so that, under the assumption that 
\begin{subequations}
	\begin{equation}
		\mathcal{P}\rightarrow \mathcal{P}+\frac{d^{(\mathcal{P})}}{dt}\mathcal{P},\qquad \mathcal{R}\rightarrow \mathcal{R}+\frac{d^{(\mathcal{R})}}{dt}\mathcal{R},
		\end{equation}
		\end{subequations} 
where $d^{(.)}/dt$ is the total derivative given by
\begin{subequations}
	\begin{equation}
		\frac{d^{(\mathcal{P})}}{dt}=\frac{\partial}{\partial t}+\dot{\zeta}\frac{\partial}{\partial\zeta},\qquad \frac{d^{(\mathcal{R})}}{dt}=\frac{\partial}{\partial t}+\dot{\beta}\frac{\partial}{\partial\beta},
		\end{equation}
		\end{subequations}
we obtain our set of coupled integro-differential flow equations for the time-dependent distribution functions $\hat{P}$ and $\hat{R}$ as
\begin{subequations}
	\label{timedep}
	\begin{equation}
		\frac{\partial}{\partial t}\mathcal{P}+\dot{\zeta}\frac{\partial}{\partial \zeta}\mathcal{P}=0,\qquad \frac{\partial}{\partial t}\mathcal{R}+\dot{\beta}\frac{\partial}{\partial \beta}\mathcal{R}=0.
	\end{equation}
\end{subequations}
They govern the RG flows of the distribution functions for the two independent random variables. Their solutions provide the required information about the time-dependent critical and off-critical flows.

To solve these equations we suppose that the distribution functions $\hat{P}$ and $\hat{R}$ are split into two terms, a leading-order term that does not contain an explicit time dependence (rather the time comes parametrically through the $\zeta(t)$, $\beta(t)$, $l(t)$ and $\Gamma(t)$ terms) and a small correction term that contains an explicit time dependence: i.e.\ $\hat{P}(t,\zeta(t),l(t);\Gamma(t))=\hat{P}^0(\zeta(t),l(t);\Gamma(t))+\hat{P}^1(t,\zeta(t),l(t);\Gamma(t))$ and $\hat{R}(t,\beta(t),l(t);\Gamma(t))=\hat{R}^0(\beta(t),l(t);\Gamma(t))+\hat{R}^1(t,\beta(t),l(t);\Gamma(t))$, with $\hat{P}^1$ and $\hat{R}^1$ small. Plugging these into Equations (\ref{timedep}) to leading order we obtain
\begin{subequations}
	\begin{eqnarray}
	\hat{P}^0_{\Gamma}-\hat{P}^0_{\zeta}-\hat{R}_0^0\hat{P}^0* \hat{P}^0-\left(\hat{P}_0^0-\hat{R}_0^0\right)\hat{P}^0&=&0,\\
	\hat{R}^0_{\Gamma}-\hat{R}^0_{\beta}-\hat{P}_0^0\hat{R}^0* \hat{R}^0-\left(\hat{R}_0^0-\hat{P}_0^0\right)\hat{R}^0&=&0,
	\end{eqnarray}
\end{subequations}
where $\hat{P}_0^0\equiv\hat{P}^0|_{\zeta(t)=0}$ and $\hat{R}_0^0\equiv\hat{R}^0|_{\beta(t)=0}$. To first order we obtain
\begin{subequations}
	\begin{eqnarray}
	\hat{P}^1_{\Gamma}-\hat{P}^1_{\zeta}-2\hat{R}_0^0\hat{P}^0* \hat{P}^1-\hat{R}_0^1\hat{P}^0* \hat{P}^0&&\nonumber\\
	\qquad\qquad-\left(\hat{P}_0^1-\hat{R}_0^1\right)\hat{P}^0-\left(\hat{P}_0^0-\hat{R}_0^0\right)\hat{P}^1&=&\hat{g}_P,\\
	\hat{R}^1_{\Gamma}-\hat{R}^1_{\beta}-2\hat{P}_0^0\hat{R}^0* \hat{R}^1-\hat{P}_0^1\hat{R}^0* \hat{R}^0&&\nonumber\\
	\qquad\qquad-\left(\hat{R}_0^1-\hat{P}_0^1\right)\hat{R}^0-\left(\hat{R}_0^0-\hat{P}_0^0\right)\hat{R}^1&=&\hat{g}_R,
	\end{eqnarray}
\end{subequations}
where $\hat{P}_0^1\equiv\hat{P}^1|_{\zeta(t)=0}$ and $\hat{R}_0^1\equiv\hat{R}^1|_{\beta(t)=0}$ and where $\hat{g}_P(\zeta)$ and $\hat{g}_R(\beta)$ are integration constants.

To find the off-critical flows for the above leading order and first-order equations, we make a Laplace transform on $\hat{P}$ and $\hat{R}$,
\begin{subequations}
	\begin{eqnarray}
		P(t,\zeta(t),y(t);\Gamma(t))&=&\int^{\infty}_0e^{-y(t)l(t)}\hat{P}(t,\zeta(t),l(t);\Gamma(t))\,dl(t),\\
		R(t,\beta(t),y(t);\Gamma(t))&=&\int^{\infty}_0e^{-y(t)l(t)}\hat{R}(t,\beta(t),l(t);\Gamma(t))\,dl(t),
	\end{eqnarray}
\end{subequations}
to get
\begin{subequations}
	\begin{eqnarray}
	P^0_{\Gamma}-P^0_{\zeta}-R_0^0P^0* P^0-\left(P_{00}^0-R_{00}^0\right)P^0&=&0,\\
	R^0_{\Gamma}-R^0_{\beta}-P_0^0R^0* R^0-\left(R_{00}^0-P_{00}^0\right)R^0&=&0,
	\end{eqnarray}
\end{subequations}
where $P_{0}^0\equiv P^0(0,y(t))$, $R_{0}^0\equiv R^0(0,y(t))$, $P_{00}^0\equiv P^0(0,0)$, $R_{00}^0\equiv R^0(0,0)$,  and
\begin{subequations}
	\label{f_order}
	\begin{eqnarray}
	P^1_{\Gamma}-P^1_{\zeta}-2R_0^0P^0* P^1-R_0^1P^0* P^0&&\nonumber\\
	\qquad\qquad-\left(P_{00}^1-R_{00}^1\right)P^0-\left(P_{00}^0-R_{00}^0\right)P^1&=&g_P,\\
	R^1_{\Gamma}-R^1_{\beta}-2P_0^0R^0* R^1-P_0^1R^0* R^0&&\nonumber\\
	\qquad\qquad-\left(R_{00}^1-P_{00}^1\right)R^0-\left(R_{00}^0-P_{00}^0\right)R^1&=&g_R,
	\end{eqnarray}
\end{subequations}
where $P_{0}^1\equiv P^1(0,y(t))$, $R_{0}^1\equiv R^1(0,y(t))$, $P_{00}^1\equiv P^1(0,0)$, $R_{00}^1\equiv R^1(0,0)$, and $g_P$ and $g_R$ are the Laplace transforms of $\hat{g}_P$ and $\hat{g}_R$ respectively. 

We can now work with the (small) inverse lengthscale $y(t)$, instead of the (large) lengthscale $l(t)$. At $y(t)=0$ we have the normalisation conditions 
\begin{equation}
	\label{norms}
		\int^{\infty}_0 P|_{y=0}\,d\zeta=1,\quad \int^{\infty}_0 R|_{y=0}\,d\beta=1.
\end{equation}
We now suppose that $P=P^0+P^1$ and $R=R^0+R^1$ and make the ansatzes \cite{fisher} that 
\begin{subequations}
	\begin{eqnarray}
	P^0=Y^0e^{-\zeta u^0},\quad P^1&=&f_P\dot{\zeta}Y^1e^{-\zeta u^1},	\\
	R^0=S^0e^{-\beta v^0},\quad R^1&=&f_R\dot{\beta}S^1e^{-\beta v^1},
\end{eqnarray}
\end{subequations}
for unknown (to be determined) functions $Y^0(y(t);\Gamma(t))$, $Y^1(y(t);\Gamma(t))$, $S^0(y(t);\Gamma(t))$, $S^1(y(t);\Gamma(t))$, $u^0(y(t);\Gamma(t))$, $u^1(y(t);\Gamma(t))$, $v^0(y(t);\Gamma(t))$, $v^1(y(t);\Gamma(t))$, $f_P(t)$ and $f_R(t)$. Denoting $Y^0(y(t)=0)\equiv Y_0^0$ and $S^0(y(t)=0)\equiv S_0^0$, to leading order we obtain 
\begin{subequations}
	\begin{eqnarray}
	\label{eq_u_0}
	u^0_{\Gamma}&=&-Y^0S^0,\\
	\label{eq_tau_0}
	v^0_{\Gamma}&=&-Y^0S^0,\\
	Y^0_{\Gamma}&=&\left(Y_0^0-S_0^0-u^0\right)Y^0,\\
	S^0_{\Gamma}&=&\left(S_0^0-Y_0^0-v^0\right)S^0.
\end{eqnarray}
\end{subequations}
The solution to these coupled equations follows directly (they closely follow the solutions presented in \cite{fisher}, however the normalisation conditions lead to a slight change) as
\begin{subequations}
	\begin{eqnarray}
	u^0(y(t);\Gamma(t))&=&-\delta(y(t))+\Delta^0(y(t))\coth\left[\left(\Gamma(t)+C(y(t))\right)\Delta^0(y(t))\right],\\
	v^0(y(t);\Gamma(t))&=&\delta(y(t))+\Delta^0(y(t))\coth\left[\left(\Gamma(t)+C(y(t))\right)\Delta^0(y(t))\right],\\
	Y^0(y(t);\Gamma(t))&=&\frac{\Delta^0(y(t))}{\sinh\left[\left(\Gamma(t)+C(y(t))\right)\Delta^0(y(t))\right]}e^{-D^0(y(t))-\left[S_0^0-Y_0^0-\delta(y(t))\right]\Gamma(t)},\\
	S^0(y(t);\Gamma(t))&=&\frac{\Delta^0(y(t))}{\sinh\left[\left(\Gamma(t)+C(y(t))\right)\Delta^0(y(t))\right]}e^{D^0(y(t))+\left[S_0^0-Y_0^0-\delta(y(t))\right]\Gamma(t)},
\end{eqnarray}
\end{subequations}
for constants of integration $\delta(y(t))$, $\Delta^0(y(t))=\sqrt{y+\delta(y(t))^2}$, $C(y(t))$ and $D^0(y(t))$.

For the first-order equations (\ref{f_order}) we make the assumptions that $u^0=u^1\equiv u$ and $v^0=v^1\equiv v$. This allows us to write down a further four coupled equations 
\begin{subequations}
	\label{offs}
	\begin{eqnarray}
		\label{eq_u_1}
	u_{\Gamma}&=&\frac{f_R}{f_P}\frac{S^1}{Y^1}\left(Y^0\right)^2-2Y^0S^0-\frac{g_P}{Y^1},\\
			\label{eq_tau_1}
	v_{\Gamma}&=&\frac{f_P}{f_R}\frac{Y^1}{S^1}\left(S^0\right)^2-2Y^0S^0-\frac{g_R}{S^1},\\
	Y^1_{\Gamma}&=&\left(Y_0^0-S_0^0-u\right)Y^1+\left(Y_0^1+\frac{f_R}{f_P}S_0^1\right)Y^0,\\
	S^1_{\Gamma}&=&\left(S_0^0-Y_0^0-v\right)S^1-\left(S_0^1+\frac{f_P}{f_R}Y_0^1\right)S^0.
\end{eqnarray}
\end{subequations}
Now, from the leading order solutions, we know that $S^0=fe^{\bar{\alpha}}$ and $Y^0=fe^{-\bar{\alpha}}$, where
\begin{subequations}
	\begin{eqnarray}
		f(y(t);\Gamma(t))&=&\frac{\Delta^0(y(t))}{\sinh\left[\left(\Gamma(t)+C(y(t))\right)\Delta^0(y(t))\right]},\\
		\bar{\alpha}(y(t);\Gamma(t))&=&{D^0(y(t))+\left[S_0^0-Y_0^0-\delta(y(t))\right]\Gamma(t)}.
	\end{eqnarray}
\end{subequations}
As such, we suppose that $S^1=ge^{\bar{\alpha}}$ and $Y^1=ge^{-\bar{\alpha}}$, for function $g(y(t);\Gamma(t))$ to be found.
To make progress we examine the form of the solution at $t=0$, and note that under our assumptions, the forms of Eq.'s (\ref{eq_u_0}) and (\ref{eq_u_1}) and Eq.'s (\ref{eq_tau_0}) and (\ref{eq_tau_1}) must correspond. 
To ensure this correspondance we have that $f_P=f_R$ and $g_P=g_R=0$. As such Eq.'s (\ref{offs}) reduce to
\begin{subequations}
	\label{f_order2}
	\begin{eqnarray}
	u_{\Gamma}&=&\frac{S^1}{Y^1}\left(Y^0\right)^2-2Y^0S^0,\\
	v_{\Gamma}&=&\frac{Y^1}{S^1}\left(S^0\right)^2-2Y^0S^0,\\
	Y^1_{\Gamma}&=&\left(Y_0^0-S_0^0-u\right)Y^1+\left(Y_0^1+S_0^1\right)Y^0,\\
	S^1_{\Gamma}&=&\left(S_0^0-Y_0^0-v\right)S^1-\left(S_0^1+Y_0^1\right)S^0.
	\end{eqnarray}
\end{subequations}
Substitution of the forms for $S^1$ and $Y^1$ and using the leading order solutions for the other functions, leads to a single second-order differential equation in $g$ that can be readily solved;
\begin{equation}
	g_{\Gamma\Gamma}-\frac{g_{\Gamma}^2}{g}=gf^2.
\end{equation}
Using the form for $f$ above we obtain
\begin{equation}
	g(y(t);\Gamma(t))=\frac{\Delta^1(y(t))e^{-D^1(y(t))\Gamma(t)}}{\sinh\left[\left(\Gamma(t)+C(y(t))\right)\Delta^0(y(t))\right]},
\end{equation}
for constants of integration $\Delta^1(y(t))$ and $D^1(y(t))$.

We are thus left with six constants of integration, namely $\delta(y(t))$, $\Delta^0(y(t))$, $C(y(t))$, $D^0(y(t))$, $\Delta^1(y(t))$ and $D^1(y(t))$. The normalisation conditions (\ref{norms}) reduce to
\begin{subequations}
	\begin{eqnarray}
		Y^0_0+f_P\dot{\zeta}Y^1_0&=&u_0,\\
		S^0_0+f_R\dot{\beta}S^1_0&=&v_0,
	\end{eqnarray}
\end{subequations}
where $u(y(t)=0)\equiv u_0$ and $v(y(t)=0)\equiv v_0$. These in turn lead to $\delta(y(t)=0)=\Delta^0(y(t)=0)$, $D^0(y(t)=0)=\Delta^0(y(t)=0)C(y(t)=0)$, $S_0^0-Y_0^0=2\delta(y(t)=0)$ and $\Delta^1(y(t)=0)=0$. It is this last condition that contains all the information on the adiabatic transfer. We recall that we began the analysis, at $t=0$, by writing down a perturbation in time to an off-critical flow (about which we know the full physics \cite{fisher}). We have then proceeded to find the form of the distribution functions at a given $t>0$. These distribution functions contain leading-order terms ($P^0$ and $R^0$) and explicit time-dependent correction terms ($P^1$ and $R^1$), that are dependent on the rates $\dot{\zeta}$, $\dot{\beta}$. 

To ensure that we remain in the vicinity of the ground state ($P^1$, $R^1$ small perturbations), we impose that the rates $\dot{\zeta}$ and $\dot{\beta}$ are small. These then become our adiabatic conditions. If at any given time (say $t=t_1$) we stop the time-flow and allow the RG scheme to take over ($y\rightarrow0$), we should expect $P^1$ and $R^1\rightarrow 0$, as we have shown above ($\Delta^1\rightarrow0$). In this case the distribution functions are simply $P=P^0$ and $R=R^0$, both evaluated at $t=t_1$, i.e.\ they are the off-critical flows in the low-energy and long-length scales that have already been determined by Fisher \cite{fisher}. This is not surprising: time acts on our Hamiltonian (Eq.\ \ref{ham}) as a simple rescaling of the site- and bond-coupling strengths.  During our adiabatic transfer we would not expect to attain these low-energy and long-length scales for all times. Those times where we do attain these scales will correspond to the already found off-critical flows, whereas those times where we do not attain these scales will correspond to a (small) error away from the lowest energy (ground) state. Therefore we expect that $\Delta^1(y(t))\sim cy$, in the limit $y$ small, for some constant $c$.


As the off-critical flows are fundamentally related to those of \cite{fisher}, we are able to state that $C(y(t))=C_0$ and $\delta(y(t))=\delta$ (both constants), and $D^0(y(t))=0$ and $D^1(y(t))=0$. Therefore for large $\Gamma$, small $y$ and small $\delta$, we obtain
\begin{subequations}
	\begin{eqnarray}
		u(y(t);\Gamma(t))&=&-\delta+\Delta^0(y(t))\coth\left[\left(\Gamma(t)+C_0\right)\Delta^0(y(t))\right],\\
		v(y(t);\Gamma(t))&=&\delta+\Delta^0(y(t))\coth\left[\left(\Gamma(t)+C_0\right)\Delta^0(y(t))\right],\\
		Y^0(y(t);\Gamma(t))&=&\frac{\Delta^0(y(t))}{\sinh\left[\left(\Gamma(t)+C_0\right)\Delta^0(y(t))\right]}e^{-\delta\Gamma(t)},\\
		S^0(y(t);\Gamma(t))&=&\frac{\Delta^0(y(t))}{\sinh\left[\left(\Gamma(t)+C_0\right)\Delta^0(y(t))\right]}e^{\delta\Gamma(t)},\\
		Y^1(y(t);\Gamma(t))&=&\frac{\Delta^1(y(t))}{\sinh\left[\left(\Gamma(t)+C_0\right)\Delta^0(y(t))\right]}e^{-\delta\Gamma(t)},\\
		S^1(y(t);\Gamma(t))&=&\frac{\Delta^1(y(t))}{\sinh\left[\left(\Gamma(t)+C_0\right)\Delta^0(y(t))\right]}e^{\delta\Gamma(t)}.
	\end{eqnarray}
\end{subequations}
In terms of our joint distribution functions $P$ and $R$, we therefore have
\begin{subequations}
	\begin{eqnarray}
		P&=& \frac{e^{-(\zeta u+\delta\Gamma)}}{\sinh\left[\left(\Gamma+C_0\right)\Delta^0\right]}\left(\Delta^0+\dot{\zeta}\Delta^1\right),\\
		R&=&\frac{e^{-(\beta v-\delta\Gamma)}}{\sinh\left[\left(\Gamma+C_0\right)\Delta^0\right]}\left(\Delta^0+\dot{\beta}\Delta^1\right).
	\end{eqnarray}
\end{subequations}
These solutions for the correction to the off-critical flows in the adiabatic time-limit (where $\dot{\zeta}$ and $\dot{\beta}$ are small) are the main results of our paper. The form for $\delta$ has been written down in Eq.\ (\ref{delta}) and provides a measure of the `distance' from the critical point.

%

\section*{III. Transition Through the Quantum Critical Point}




Up to now we have concentrated on the ground, or equilibrium, state properties of the time-dependent transverse-field Ising model, using the time as a weighting on the random bond and site couplings. The inclusion of a time in the RG analysis provides the equilibrium critical exponents such as the location of the critical point, magnetisation and correlation length scalings, all parametrised in time. In this section we are interested in the non-equilibrium dynamics that occur as we time-evolve the governing Hamiltonian (Eq.\ (\ref{ham})) through a critical point. As noted earlier, our choice of an adiabatic time evolution is deliberately made in order to remain in the local (in-time) ground state throughout. We thus do not expect any defects - or excited states - to be generated, except in the vicinity of the critical point where the energy gap between the ground state and first excited state will decrease and disappear exactly at the critical point. This is precisely where the computation loses adiabaticity. 

It turns out that the analytic scaling for the defect density (a non-equilibrium dynamics) is dependent on knowledge of the equilibrium critical exponents \cite{dz2,rev1}. We will assume for the transition dynamics that we can separate out the time functions from the random variables. Thus we can, by using the equilibrium scalings for the correlation lengths $\xi(t)$, the distance from the critical point $\delta$, and the form of the time functions $A_{ij}(t)$ and $B_i(t)$, find an estimate for the density of defects (or the number of domain walls) that are formed as we transfer through this non-adiabatic region.

To begin, we therefore consider a scaling analysis using the forms of the critical parameter $\delta$ (Eq.\ (\ref{delta})), critical correlation length $\xi$, given by
\begin{equation}
	\label{crit_corr}
	\xi=\frac{1}{|\delta|^{\nu}},
\end{equation}
with $\nu=2$, and the dynamical $z$ exponent, given by
\begin{equation}
	\label{z_exp}
	z\sim\frac{1}{|\delta|+\Delta^0},
	\end{equation}
all in the region of the critical point. The energy gap is defined as $\Delta_{\text{gap}}\sim|\delta|^{z\nu}\sim|\delta|^{2/[|\delta|+\Delta^0]}$. We are interested in the region close to the critical point, specifically the region at which the computation loses adiabaticity. This occurs at a critical $\hat{\delta}$ when the energy gap $\Delta_{\text{gap}}$ is equal to the rate of approach towards the critical point, defined as $\Delta_{\text{rate}}$. To find $\Delta_{\text{rate}}$ we note that we must separately take account of the rate of transition of both of the coupling strengths, $J_{ij}(t)$ and $h_i(t)$. From these, we expect to be able to write down a characteristic rate $1/\tau$ as
\begin{equation}
	\frac{1}{\tau}=\frac{1}{\tau_{{A}(t)}}+\frac{1}{\tau_{{B}(t)}},
	\end{equation}
where $1/\tau_{{A}(t)}$ and $1/\tau_{{B}(t)}$ represent the characteristic rates of the evolution of the bond and site couplings, respectively. These rates are found through the derivatives of the time functions ${A}(t)$ and ${B}(t)$ evaluated at $t_f$, the time when the system passes through the critical point. Then $\tau_{{A}(t_f)}=1/{|\dot{A}(t_f)|}$ and $\tau_{{B}(t_f)}= 1/{|\dot{B}(t_f)|}$. For example, if ${A}(t)=(t/T)^2$ and ${B}(t)=1-(t/T)^2$ (as in Fig.\ \ref{schem}, where we assume $\Delta_{\tilde{h}}=\Delta_{\tilde{J}}$), then $t_f = T/\sqrt{2}$ and $\tau_{{A}(t_f)}=\tau_{{B}(t_f)}=T/\sqrt{2}$ and $\tau=T/(2\sqrt{2})$. 
This characteristic rate $1/\tau$ introduces a natural energy scale on the adiabatic evolution of the RG scheme: we can define a critical $\Gamma$, defined by $\Gamma_{\tau}\sim\ln(\Omega\tau)$. We must impose that $\Gamma<\Gamma_{\tau}$ in order for the RG scheme to be valid in the evolution of the system.

Close to the critical point $\delta$ can be also linearised in time, such that $\delta(t_f)\sim t_f /\tau$, which gives $\Delta_{\text{rate}}=(\tau\delta)^{-1}$. Thus, we can now find $\delta_c$ as the solution to
\begin{equation}
	\frac{a}{\tau\delta_c}={\delta}_c^{{2/[|{\delta}_c|+\Delta^0]}},
\end{equation}
where $a$ is a free parameter. Now, $\Delta^0=\sqrt{y+{\delta}_c^2}$, so we are left with two options: either we expand about small ${\delta}_c$, or we expand about small $y$. In the first case, using the result for ${\delta}_c$, together with the critical correlation length (Eq.\ (\ref{crit_corr})), we find that ${\xi}_c\equiv\xi|_{{\delta}_c}$ approximates to
\begin{equation}
	\label{xx}
	{\xi}_c\approx\left(\frac{\log\left[\log({\tau/a})\right]-\log({\tau/a})}{\sqrt{y}\log({\tau/a})-\left(\sqrt{y}+2\right)\log\left[\log({\tau/a})\right]}\right)^2\approx\frac{\log^2\left(\tau/a\right)}{\log^2\left[\log({\tau/a})\right]},
	\end{equation}
since $y$ is small, and therefore can be safely neglected. In the second case when we expand about small $y$, we have $\Delta^0\sim{\delta}_c$, in which case we revert back to the analysis of \cite{dz2,italians}, the result of which is identical to the right hand side of Eq.\ (\ref{xx}). It is important to note that the choice of functions $A_{ij}(t)$ and $B_i(t)$ enters the above result through the characteristic rate $1/\tau$; in particular we can take $A_{ij}(t)$ and $B_i(t)$ to be nonlinear functions of $t/T$, as has been considered in the non-random transverse-field Ising model \cite{nonlinear1,nonlinear2,nonlinear3}.
	
For this adiabatic transition through the critical point, the logarithmic scaling reflects the departure from the typical Kibble-Zurek ($\propto \tau^{1/2}$) scaling that is evidenced in a pure (non-random), or weakly disordered, Ising model. This is the same scaling as found analytically by \cite{dz2} and numerically by \cite{italians}, however our analysis, through the inclusion of the time-dependent parameters $A_{ij}(t)$ and $B_i(t)$ into the governing Hamiltonian, has established a generic characteristic timescale that represents a transition through the critical point that is influenced by {\it{both}} the bond strengths, $J_{ij}(t)$, and the site strengths, $h_i(t)$. \\

\section*{IV. Discussion and Conclusion}

In this work, we have developed and extended the well-established real-space RG approach for quantum chains with randomness  \cite{fisher, fisher92, fisher94} to the case of time-dependent Hamiltonians. We took the quantum Ising chain with random terms as an example to illustrate the protocol of the RG approach. In every step, the parallels and differences from the time-independent case have been noted. 

In particular, the location of the quantum critical point, or points, in time depend on the functions that carry the time dependence in the Hamiltonian. This time interval is defined by the time period between the zero of the exchange interaction term in the ARQIM and the zero of the transverse field term.
The off-critical flows have been calculated and the ground state properties of the ARQIM have been investigated. In general, the magnetisation as well as the correlation functions retain the form obtained by Fisher \cite{fisher, fisher92, fisher94} and we have indicated in detail where the time enters the calculation.

An important outcome is the connection with the Kibble-Zurek dynamics which govern the system near the critical point. We have constructed a scaling argument for the density of defects as we adiabatically pass through the critical point of the system, in either direction. Our method, thus, provides a justification of the exponents and connects the Kibble-Zurek dynamics with the microscopic time-dependent Hamiltonian.

A further consequence of our calculations is the ability to modify (or shift in time) the location of the quantum critical point. Such a feature can be of great importance to experiment: one can imagine that a device could be constructed to simulate the quantum critical point. A possible device in this regard would be the development of a long 1D flux qubit chain. Reference \cite{az_1} has developed a prototype qubit-coupler-qubit device where the governing Hamiltonian is given by Eq.\ (\ref{ham}) for $N=2$. In particular the site and bond coupling terms can be assumed to be independent and random, and crucially individually  modifiable. In the future we would expect that the length of the qubit chain could be increased to a few thousand qubits. This would put experiment in the $N$-large limit where the conclusions of this paper could be tested.

The RG approach that we introduce in this paper can be further extended to consider non-integrable spin-chain models, such as the XXZ or next-nearest neighbour Hamiltonians \cite{vosk_altman,xxz,pekker,v_huse_a} with random interactions. These out-of-equilibrium models admit a many-body localization phase transition that is currently the focus of much interest (see for example \cite{v_huse_a,zhang}). The inclusion of time-dependent functions into these Hamiltonians, along the lines of the adiabatic functions $A(t)$ and $B(t)$ considered in this paper, will allow us to systematically probe (some of the) properties of this transition in further detail. For example, we expect to be able to understand the identified fractal thermal Griffiths regions \cite{zhang}. 

The quantity that will enable the characterization of different phases in many different settings is the entanglement entropy. In simpler cases \cite{cole},  such as out-of-equilibrium steady states of the quantum Ising model, or when perturbations break the integrability, critical boundaries were characterised by the value of the central charge. Following Ref.'s \cite{RefaelMoore,Devakul}, we expect to be able to calculate the distribution of the entanglement entropy for the ARQIM using the real-space RG technique developed in this paper,
but in non-integrable spin models, how the time-dependence affects the entanglement entropy especially in connection to many-body localization, is of prime interest and an open question in this context.

%

\section*{Acknowledgments}

We would like to thank Alexander Balanov, Claudio Castelnovo, John Chalker, Andrey Chubukov, Mike Gunn, Vedika Khemani, David Pekker, Anatoli Polkovnikov and Artur Sowa for insightful discussions that took place during this work. The work was supported by EPSRC through the grant EP/M006581/1.

\section*{Appendix}

In this appendix we provide further details regarding the renormalization-group analysis of Sect.\ II. We start with Eq.\ (\ref{ham}), noting that in the first case when the largest coupling is a site, we have $\Omega(t)=h_{i}(t)$ for some $i$. The local Hamiltonian for this site and the two adjoining bonds is then $H^s_{i-1,i+1}(t)=-h_i(t)\sigma_i^x-J_{i-1,i}(t)\sigma_{i-1}^z\sigma_i^z-J_{i,i+1}(t)\sigma_i^z\sigma_{i+1}^z$, for  adjoining bonds $J_{i-1,i}(t)$ and $J_{i,i+1}(t)$. Under the RG procedure, these adjoining bonds can be treated perturbatively resulting in an effective Hamiltonian for $H^s_{i,i+1}(t)$, written as
\begin{subequations}
\begin{equation}
	\hat{H}^s_{i-1,i+1}(t)\approx-\hat{J}_{i-1,i+1}(t)\sigma_{i-1}^z\sigma_{i+1}^z,
\end{equation}
where 
\begin{equation}
	\hat{J}_{i-1,i+1}(t)=\frac{J_{i-1,i}(t)J_{i,i+1}(t)}{h_i(t)}.
\end{equation}	
\end{subequations}
This is site decimation: a site, $h_i(t)$, and the two adjoining bonds, $J_{i-1,i}(t)$ and $J_{i,i+1}(t)$, are removed and a new bond, $\hat{J}_{i-1,i+1}(t)$ created, joining sites $h_{i-1}(t)$ and $h_{i+1}(t)$. The new bond lengthscale under this decimation becomes $\hat{l}^B_{i-1,i+1}={l}^B_{i-1,i}+{l}^B_{i,i+1}+{l}^S_{i}$.  As in \cite{fisher}, we set each initial ${l}^S_{i}$ and ${l}^B_{i,i+1}$ equal to 1/2.

In the second case when the largest coupling is a bond, we have $\Omega(t)=J_{i,i+1}(t)$ for some $i$. The local Hamiltonian for this bond and two adjoining sites is then $H^b_{i,i+1}(t)=-h_i(t)\sigma_i^x-h_{i+1}(t)\sigma_{i+1}^x-J_{i,i+1}(t)\sigma_{i}^z\sigma_{i+1}^z$, for adjoining sites $h_{i}(t)$ and $h_{i+1}(t)$. Under the RG procedure, these adjoining sites can be treated similarly, resulting in an effective Hamiltonian for $H^b_{i,i+1}(t)$, written as
\begin{subequations}
\begin{equation}
	\hat{H}^b_{i,i+1}(t)\approx-\hat{h}_{i}(t)\sigma_{i}^x,
\end{equation}
where 
\begin{equation}
	\hat{h}_{i}(t)=\frac{h_{i}(t)h_{i+1}(t)}{J_{i,i+1}(t)}.
\end{equation}	
\end{subequations}
This is bond decimation: a bond, $J_{i,i+1}(t)$, and the two adjoining sites, $h_{i}(t)$ and $h_{i+1}(t)$, are removed and a new spin cluster, $\hat{h}_{i}(t)$ created, with adjoining bonds $J_{i-1,i}(t)$ and $J_{i+1,i+2}(t)$. The new spin cluster lengthscale under this decimation becomes $\hat{l}^S_{i}={l}^S_{i}+{l}^S_{i+1}+{l}^B_{i,i+1}$. 

The above perturbation analysis formally requires that $\hat{J}_{i-1,i+1}(t)$ is small ($<\epsilon$) for a bond decimation and that $\hat{h}_{i}(t)$ is small for a site decimation. However, this stipulation can be somewhat relaxed: errors that result in the initial few steps of the renormalization from invalidity of the perturbation are gradually smoothed away as the number of RG steps increases such that the RG flows become asymptotically valid in the low-energy, long-distance limit.

In the case when we separate out the time functions the criteria for validity of the RG scheme becomes slightly more intricate. 
To see this, first note that the above expressions for 
$\hat{J}_{i-1,i+1}(t)$ and $\hat{h}_i(t)$ refer to the distribution functions after a single renormalization step of either bond or site decimation. In general, and after many (bond and site) renormalization steps, a site decimation will give the $\hat{J}_{i-1,i+1}(t)$ as (explicitly separating out the time dependence):
\begin{subequations}
	\label{conds}
\begin{equation}
	\label{conds_site}
	\hat{J}_{i-1,i+\hat{l}-1/2}(t)=\frac{\prod_{k=i-1}^{k=i+\hat{l}-3/2}A_{k,k+1}(t)\tilde{J}_{k,k+1}}{\prod_{k=i}^{k=i+\hat{l}-3/2}B_k(t)\tilde{h}_k},
\end{equation}
while a bond decimation will give the $\hat{h}_i(t)$ as
\begin{equation}
	\label{conds_bond}
		\hat{h}_{i}(t)=\frac{\prod_{k=i}^{k=i+\hat{l}-1/2}B_k(t)\tilde{h}_{k}}{\prod_{k=i}^{k=i+\hat{l}-3/2}A_{k,k+1}(t)\tilde{J}_{k,k+1}}
	\end{equation}
\end{subequations}
where $\hat{l}=n+1/2$, for integer $n$, provides the bond or spin cluster length. 

While we are able to absorb possible errors during the decimation steps, to end up in the low-energy state, and hence for validity of the RG approach, we must impose that, after many RG steps, a site decimation leads to small  $\hat{J}_{i-1,i+\hat{l}-1/2}(t)$ (Eq.\ (\ref{conds_site})), while a bond decimation satisfies small $\hat{h}_{i}(t)$ (Eq.\ (\ref{conds_bond})). We must therefore consider the ratio of the time-dependent functions $A_{ij}(t)$ and $B_i(t)$ together with the ratio of the time-independent random distributions. 
However, these validity arguments must be placed into context with regards to the maximum coupling, as defined in $\Omega(t)=\max_{ij}\left(J_{ij}(t),h_i(t)\right)$. 
As such, the cases $A(t)\ll B(t)$ or $A(t)\gg B(t)$ will impose some preference for site or bond decimation, respectively. This indicates that, for $A(t)\ll B(t)$, we would expect a predominance of site decimations while similarly, for $A(t)\gg B(t)$, we would expect a predominance of bond decimations. 

\end{document}